\documentclass[aps,apl,twocolumn,superscriptaddress]{revtex4}
\usepackage{graphicx}
\usepackage{dcolumn}
\usepackage{bm}
\usepackage{amsmath}

\begin{document}
\DeclareGraphicsExtensions{.eps}


\title{Stochastic optimization of a cold atom experiment using a genetic algorithm}

\author{W. Rohringer}
 \affiliation{Atominstitut der \"Osterreichischen Universit\"aten, TU-Wien, Stadionallee 2, 1020 Vienna, Austria}
\author{R. B\"ucker}
 \affiliation{Atominstitut der \"Osterreichischen Universit\"aten, TU-Wien, Stadionallee 2, 1020 Vienna, Austria}
\author{S. Manz}
\affiliation{Atominstitut der \"Osterreichischen Universit\"aten, TU-Wien, Stadionallee 2, 1020 Vienna, Austria}
\author{T. Betz}
\affiliation{Atominstitut der \"Osterreichischen Universit\"aten, TU-Wien, Stadionallee 2, 1020 Vienna, Austria}
\author{Ch. Koller}
 \affiliation{Atominstitut der \"Osterreichischen Universit\"aten, TU-Wien, Stadionallee 2, 1020 Vienna, Austria}
\author{M. G\"obel}
 \affiliation{Atominstitut der \"Osterreichischen Universit\"aten, TU-Wien, Stadionallee 2, 1020 Vienna, Austria}
\author{A. Perrin}
 \affiliation{Atominstitut der \"Osterreichischen Universit\"aten, TU-Wien, Stadionallee 2, 1020 Vienna, Austria}
 \author{J. Schmiedmayer}
 \affiliation{Atominstitut der \"Osterreichischen Universit\"aten, TU-Wien, Stadionallee 2, 1020 Vienna, Austria}
 \author{T. Schumm}
 \email{schumm@atomchip.org}
\affiliation{Atominstitut der \"Osterreichischen Universit\"aten, TU-Wien, Stadionallee 2, 1020 Vienna, Austria}


\date{\today}

\begin{abstract}
We employ an evolutionary algorithm to automatically optimize different stages of a cold atom experiment without human intervention. This approach closes the loop between computer based experimental control systems and automatic real time analysis and can be applied to a wide range of experimental situations. The genetic algorithm quickly and reliably converges to the most performing parameter set independent of the starting population. Especially in many-dimensional or connected parameter spaces the automatic optimization outperforms a manual search. \end{abstract}


\maketitle

In experimental research, scientists are often confronted with the task of finding ideal parameters to perform a measurement. Although in modern setups, experimental parameters are handled by computer-based control systems and also the evaluation of the experimental output is often performed by real-time computer systems, optimization is still essentially carried out by hand. 

\begin{figure}
\includegraphics{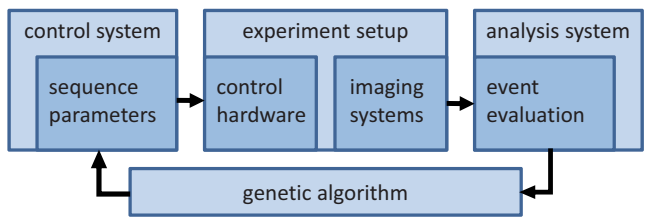}
\caption{The feedback loop between control system, experiment and real-time analysis is closed by the genetic algorithm.
\label{fig1}}
\end{figure}

In this article we present the application of a genetic algorithm to automatically optimize control parameters in an atomic physics experiment. Such algorithms have been employed to tackle optimization problems in different fields, ranging from mathematics and algorithmics~\cite{Anderson05} over economics \cite{Burke99} and aerospace engineering~\cite{Obayashi00} to game theory~\cite{Weibull95}. Our algorithm generates a set of random starting parameters, automatically performs the experiment and recovers information about the performance of the parameter set from real-time analysis of the experimental result. This information is used to generate a next generation of parameters with better performance, until an optimum is reached (see figure\,\ref{fig1}). This approach can be implemented in any experimental setup that requires optimization, provided that it is computer controlled and features real-time analysis.

The experimental setup used in these investigations is a single-chamber atom chip setup for the creation and study of Bose-Einstein condensates (BEC)~\cite{Wildermuth2004}. Pre-cooled $^{87}$Rb atoms are loaded into a magnetic trap created by current carrying macroscopic copper structures. Evaporative cooling is applied, creating a sample of $~10^{6}$ atoms at microkelvin temperatures. These atoms are loaded onto an atom chip, where microscopic traps are generated by lithographically patterned wires~\cite{Groth04,Trinker2008a}, and cooled to quantum degeneracy in a second phase of evaporative cooling. The wire structures on the atom chip allows a large variety of experiments like matter wave interferometry~\cite{Schumm2005} or the study of low-dimensional quantum gases~\cite{Hofferberth2007}. After the experiment is performed, the atomic density distribution is recorded using standard absorption or fluorescence imaging~\cite{Ketterle99}, either in the atom chip trap or in free time-of-flight expansion. As the atomic sample is destroyed in the imaging process, the above procedure has to be repeated many times when parameters are changed and to accumulate statistics. The duration of an experiment is 35\,s, the experimental cycle is repeated continuously and data taking can cover several days. It shall be pointed out that individual operations within the experimental sequence take place on a timescale on the order of tens of microseconds, necessitating a relative temporal resolution on the $10^{-6}$ level. A commercial real time control system (ADwin Pro I) in connection with a custom interface software coordinates the 60 involved individual devices. They are governed by a total parameter space of 2300 values during a typical experimental sequence, a subspace of which is made accessible for each individual optimization task.

\begin{figure}
\includegraphics{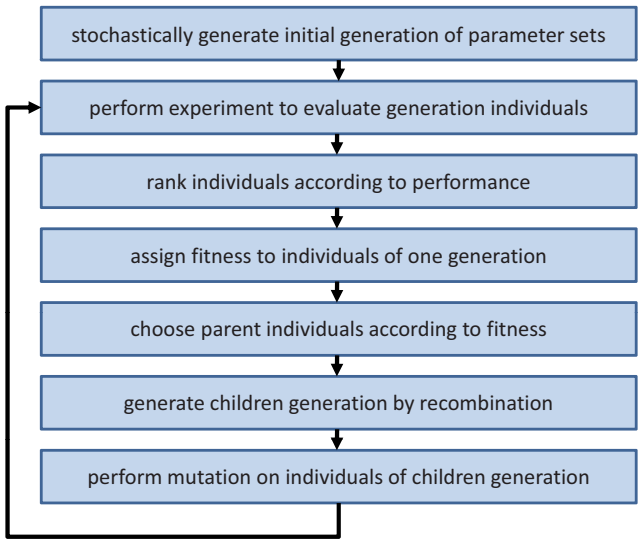}
\caption{Block diagram of the optimization process.
\label{fig2}}
\end{figure}

\begin{figure*}
\includegraphics{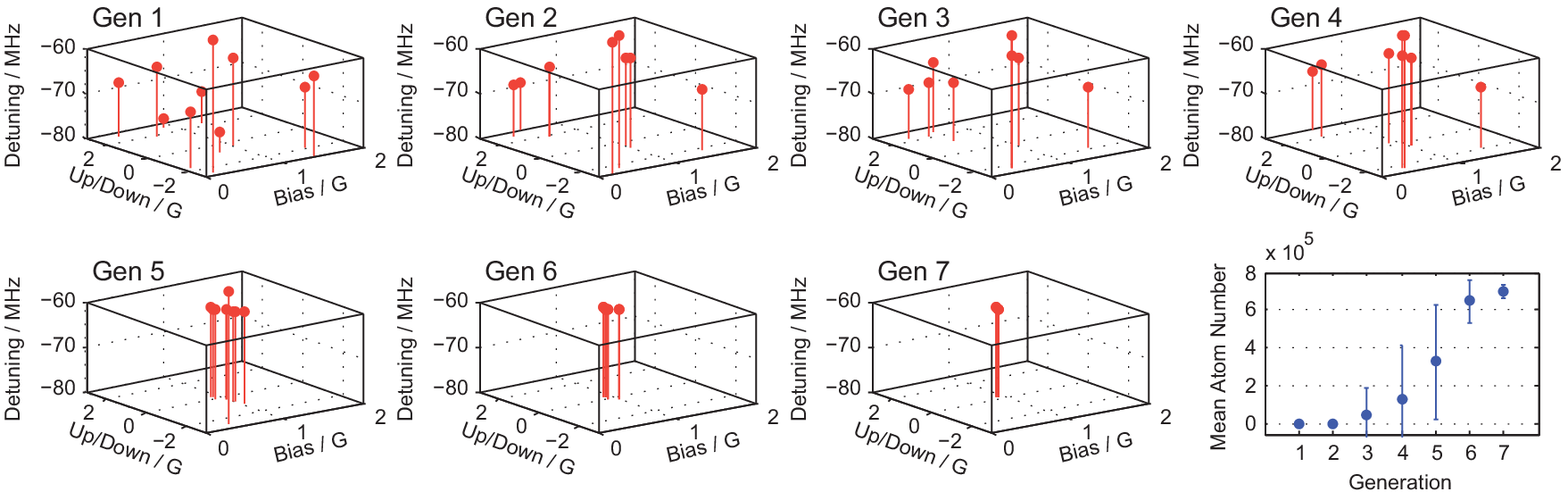}
\caption{Positions of the individuals in the three-dimensional parameter space through seven generations. Convergence is clearly observed. On the bottom right, the achieved mean atom number for each generation is depicted. The error bars correspond to the standard deviation of the individual's results within each generation. 
\label{fig3}}
\end{figure*}

Each experimental cycle produces an image of the atom cloud which yields a projected 2D measurement of the particle density distribution. Important physical quantities like the total number of atoms $N$, the peak density $n_0$, the position of the cloud or the phase space density can be derived already from a single picture. In terms of optimization theory, the above quantities can be interpreted as objective functions $f(j_i)$, where $j_i$ denotes a single parameter set out of the overall parameter space. Optimizing the experimental setup corresponds to finding a parameter set $j_i$ that maximizes (minimizes) a given objective function. Probing this function hence means performing the experimental sequence with a given parameter set and determining the desired measurement value. Since in our experiments usually only few or even a single output quantity is considered, the problem is a prime example of a single-objective optimization problem. Such tasks can be tackled effectively by stochastic algorithms~\cite{Assion98}.

The simplest approach would be to evaluate the objective function on the entire parameter space. However, the number of necessary experimental runs grows rapidly with the dimensions of the parameter space. With 10 evaluations per parameter, a scan over 1, 2, 3, 4 parameters (i.e. 10, 100, 1000, 10000 parameter sets) takes about 6~min, 1~h, 10~h or 4~days respectively.

An alternative approach are deterministic algorithms such as hill climbing~\cite{Weise08}, where knowledge on the gradient of the objective function is used to generate improved parameter sets. They perform well for problems that behave approximately linear in the objective function and guarantee convergence for unimodal problems represented by objective functions featuring only one extremum. For multimodal problems, however, they are prone to converging against local extrema.

Stochastic algorithms trade guaranteed convergence for shorter runtime in applications with a high number of (interacting) parameters~\cite{Weise08}. They introduce a random element into the selection of points in parameter space, there is no formal way to prove whether a found solution is really the global optimum. Within stochastic algorithms, genetic algorithms belong to the class of parallel optimization algorithms which do not rest on the evaluation of individual states in parameter space but rather evaluate the performance of an ensemble of states (generation). Compared with local algorithms, this approach is less dependent on the specific initial conditions for each optimization run. Out of the different parallel methods, genetic algorithms are well explored, with numerous known implementational strategies~\cite{Baricelli62,Baricelli63}.

Figure 2 illustrates the working concept of our genetic algorithm. In a first step, a starting population is distributed randomly in a parameter sub-space spanned by 2-5 variables associated with a certain stage of the experiment. Sometimes population individuals near a known optimum or on the edges of parameter space are added to speed up convergence or test the performance of the algorithm. The size of the initial population has to be chosen carefully: Large populations are time consuming as they increase the number of evaluations per generations as well as the number of generations. Too small starting populations have shown to lead to premature convergence as the diversity of the individuals decreases quickly. An intermediate initial population size of 10-15 individuals has shown to perform well in various optimization tasks. 

To evaluate the performance of a population, the experimental sequence is executed with the parameter settings represented by the individuals. The individuals are then sorted according to their performance with respect to the selected objective function, e.g. particle number or temperature. 

In analogy to evolution terminology, the fitness of a population individual F($j_{i}$) is a measure for the probability for this state to be selected for reproduction. We have chosen linear ranking based fitness assignment~\cite{Baker87}, which considers the relative performance of individuals but is independent of the achieved absolute value of the objective function. It distributes the individual fitnesses between 0 and 1: $F(j_{i})=\frac{2}{n}(1-\frac{i-1}{n-1})$ where $n$ is the size of the population.

The selection process determines the parent individuals that will 'breed' the next generation. From several possible selection schemes we have chosen stochastic universal sampling~\cite{Baricelli62} where parent individuals are chosen according to their fitness but which still introduces a stochastic process so that less performing individuals can contribute to the next generation and population diversity is kept up. The total number of parent states is adjusted to 2/3 of the total population with 2 children per pair of parent states. Since the population size is chosen to be fixed throughout the optimization process, the free slots are filled up by the best states from the previous generation. This makes our algorithm 'elitist', assuring that the states showing best performance are not lost during the optimization process, speeding up convergence.

The next generation individuals are created by intermediary recombination~\cite{Pohlheim99, Herrera98}. The parameters of the two parent states are taken as limiting edges of a hypercube in parameter space, the two child states are distributed randomly within this volume. To avoid too rapid decrease of the children's parameter space, its volume is stretched by a factor 1.5~\cite{Pohlheim99, Herrera98}.

Mutation is introduced to further impede convergence against local extrema. Its influence on the performance of genetic algorithms has been investigated in great detail and several well performing schemes are known~\cite{Haupt00, Herrera98}. For each state within one generation, there is a certain probability for random alteration, in our case 1~\%, with the additional constraint that only one of the state's variables is affected. The mutated values are obtained by a mechanism shown in~\cite{Muehlenbein95}, that constructs $j_i^{mut}=j_i+s_i\cdot rD_i\cdot 2^{-uk}$, where $j_i^{mut}$ and $j_i$ denote the mutated and the source state respectively, $s_i$ randomly choses the sign of the mutation step and $r$ defines the mutation range as a fraction of the accessible parameter space $D_i$. The last factor designates the actual distribution, characterized by the random number $u$ uniformly distributed in $[0,1]$ and the mutation precision $k$ (M\"uhlenbein's mutation). With $r=0.1$ and $k=10$ we have adapted established values~\cite{Pohlheim99}. After the mutation step the next child population is ready to be tested on the experiment.

For simplicity, the optimization cycle is repeated for a preset number of 10 times or is stopped manually when the diversity in the population has dropped to the point that all states are equal. More evolved convergence criteria like the distance of individuals in parameter space or convergence of the objective function can be implemented easily. 

The described genetic algorithm has been used to optimize various stages of the experimental sequence \cite{Wildermuth2004} like the optical pumping into magnetically trappable states or the evaporative cooling ramps. Figure~3 shows an example of an optimization of critical experimental parameters during the phase of optical molasses, which effectuates an additional cooling of the atoms coming from the magneto-optical trap before they are loaded into a magnetic wire trap. The three-dimensional parameter space is spanned by two magnetic fields (Up/Down and Bias), and the detuning of the cooling laser frequency with respect to the atomic transition. The two magnetic fields have to be optimized to compensate the earth's magnetic field and possible unknown stray fields, the laser frequency has to be optimized to achieve an efficient cooling (high detuning) while keeping the atoms confined (low detuning). As shown in figure~3 the genetic algorithm converges to optimum parameters within 7~generations with a starting population of 10 individuals. The overall runtime was 30~minutes, as compared to 386~h it would need to sample the entire parameter volume with the same resolution. Note that the optimum was found although the first two generations contained not a single well performing individual.
The algorithm rapidly and reliably finds optimum values for various optimization tasks and reproduces or outperforms manually optimized parameters sets. For too small starting populations, it has shown to be prone to premature convergence leading to sub-optimal final parameters. Numerous characteristics of the algorithm such as the size of the starting population, the amount of reproducing individuals per generation or the mutation rate and range can be adapted, weighing optimality against runtime.

In future work we will extend the approach to objective functions that cannot be evaluated in a single experimental run (image) but probe temporal dynamics of the system or the scattering of measured parameters around the mean value, requiring several measurements to evaluate a single parameter set. This will allow applications of the genetic algorithms beyond mere optimization tasks, like the reduction of excitations or the enhancement of squeezing or entanglement in dynamic matter wave interferometry~\cite{Esteve2008}. Ultimately, we aim to develop "experimental optimal control", in analogy to optimal control theory, where the result of the optimization directly provides scientific insight, not accessible by other means.

To conclude, we have implemented a robust and efficient stochastic genetic algorithm to automatically perform time-consuming optimization tasks in a complex cold atom experiment. In the given example, the algorithm converged to optimal settings within 30 min, as compared to several hours typically required for manual search. Our approach closes the loop between computer-based experimental control systems and real-time data analysis and can be implemented in many situations in experimental research.

This work was supported by the EU Integrated Project FET/QUIP "SCALA" and the Austrian Science Fund FWF Projects P20372 and P21080.


\end{document}